\documentclass[11pt]{article}

\usepackage[T1]{fontenc}
\usepackage[utf8]{inputenc}
\usepackage{lmodern}
\usepackage[margin=1in]{geometry}
\usepackage{microtype}
\usepackage{amsmath,amssymb,amsthm,mathtools}
\usepackage{physics}
\usepackage{cite}
\usepackage{tikz}
\usepackage{booktabs}
\usepackage{enumitem}
\usepackage[hidelinks]{hyperref}
\usepackage[nameinlink,capitalize]{cleveref}
\usepackage{xcolor}
\usepackage{graphicx}
\usepackage{listings}
\usepackage{framed}

\theoremstyle{definition}

\usepackage[many]{tcolorbox}
\tcbset{colback=gray!3,colframe=gray!50!black,boxrule=0.2pt,arc=1mm}

\newcommand{\cO}{\mathcal{O}}

\newcommand{\vd}{\vec d}
\newcommand{\ve}{\vec e}

\newcommand{\sst}{{s_\ast}}

\usepackage{xcolor}

\newlength{\myboxbeforeskip} 
\newlength{\myboxafterskip}  
\newlength{\myboxframesep}   
\newlength{\myboxframerule}  

 \numberwithin{equation}{section}

\usepackage[T1]{fontenc}
\usepackage[utf8]{inputenc}
\usepackage[svgnames, table]{xcolor}
\colorlet{framecolor}{Gold}
\colorlet{backcolor}{Orange}

\usepackage{eqparbox}
\usepackage{makecell}
\usepackage{bigstrut}

\begin{document}

\begin{titlepage}
\begin{center}
\vspace*{1.5cm}
{\LARGE\bfseries GLSMs for non-polynomial deformations and\\[4pt]
generalized complete intersection Calabi-Yau manifolds\par}
\vspace{1.2cm}
{\large Lara B.~Anderson\footnote{lara.anderson@vt.edu},
James Gray\footnote{jamesgray@vt.edu},
Eric Sharpe\footnote{ersharpe@vt.edu}, and
Albert Haochen Zhang\footnote{alberthj25@vt.edu}\par}
\vspace{0.6cm}
{\itshape Department of Physics, Robeson Hall 0435, Virginia Tech,\\
850 West Campus Drive, Blacksburg, VA 24061, U.S.A.\par}
\vspace{1.5cm}
\end{center}

\begin{abstract}
\noindent
Gauged Linear Sigma Models (GLSMs) have long been a key tool in understanding Calabi-Yau compactifications of string theory. While descriptions of complete intersections of polynomial defining equations have long been understood in this context, a full treatment of non-polynomial complex structure deformations has been absent from the literature. Likewise, the somewhat more exotic generalized complete intersection Calabi-Yau construction has resisted a GLSM description. In this paper we resolve both of these omissions using completely standard GLSMs with polynomial superpotentials. The non-polynomial structure seen in the geometric constructions only arises in the associated GLSM upon restriction to the vacuum space of the theory. This feature enables the construction to avoid singularities in field space that could have arisen if non-polynomial structure were incorporated directly into  the GLSM superpotential.

\end{abstract}
\vfill
\end{titlepage}
\pagebreak
\tableofcontents

\newpage
\section{Introduction}

Gauged Linear Sigma Models (GLSMs) have long been an invaluable tool in understanding Calabi-Yau compactifications of string theory \cite{Witten:1993yc}. They provide a relatively simple field theory, in which many computations can be carried out, that flows in the infrared (in an appropriate phase) to the much more complicated non-linear sigma model associated to the Calabi-Yau manifold. An example of the utility of GLSMs is that their phase structure can relate different geometries, as well as CFTs without a geometric interpretation.

Two gaps in the literature to date have been the lack of an understanding of how to describe finite non-polynomial complex structure deformations in GLSMs (see e.g.~\cite{Candelas:1987kf,Green:1987rw,Berglund:1994my,Batyrev:1993oya,Adams:2023imc}), and the closely related question of how to describe generalized complete intersection (gCICY) manifolds (first described in \cite{Anderson:2015iia} and further studied in e.g.~\cite{Berglund:2016yqo,Berglund:2016nvh,Berglund:2022dgb,Garbagnati:2017rtb,Jia:2018iza,Anderson:2015yzz,Larfors:2020weh,Kanazawa:2026wqh,Kanazawa:2026zzr}) as GLSMs (see e.g.~\cite{Berglund:2016nvh} for efforts in this direction). In this paper, we address both of these omissions.

To understand the difficulty in describing non-polynomial complex structure deformations in GLSMs, it is important to note that
there are two types of such deformations for complete intersection Calabi-Yau manifolds. The first is the restriction of complex structure deformations of the ambient space to the Calabi-Yau manifold. These have been studied in the GLSM context, for infinitesimal deformations, in \cite{Adams:2023imc}. The second arises from contributions to defining relations that are holomorphic sections of the normal bundle on the Calabi-Yau manifold, but which are not associated to holomorphic sections of the normal bundle on the ambient space. In other words, these are sections of the normal bundle on the Calabi-Yau manifold that are not polynomials of ambient space coordinates.  This second type of non-polynomial deformation can be described, as we will review, in terms of rational functions of ambient space coordinates. Because GLSM fields are typically associated to ambient space coordinates, it has been unclear how to describe such deformations in this context, as these rational functions naively lead to singularities somewhere in field space.

Similar issues arise when trying to construct GLSMs for generalized complete intersection Calabi-Yau (gCICY) manifolds  \cite{Anderson:2015iia}. In a normal complete intersection one defines a Calabi-Yau manifold as the common solution set of a number of equations that are polynomial in the ambient space coordinates. To construct a gCICY, one begins with an ordinary (non-Calabi-Yau) complete intersection ${\cal M}$, then restricts further to the vanishing locus of 
sections of line bundles which have no sections on the ambient space, but do have sections when restricted to ${\cal M}$. These extra defining relations, which describe a Calabi-Yau locus inside ${\cal M}$, can be described by rational functions in the coordinates of the ambient space. One way to think about this is that gCICYs are examples exhibiting the limit where {\it all} of the complex structure associated to some defining relations are non-polynomial.  In particular, one then immediately encounters the same issues in GLSM constructions as for non-polynomial deformations.

Our construction of GLSMs for such non-polynomial deformations and gCICYs can be outlined as follows. We enhance the field space of the GLSM by adding additional coordinates, $u_{\alpha}$, beyond those associated to the ambient space of the geometric construction. We add one such coordinate for each generator of the desired non-polynomial deformations or gCICY defining relations. We also modify the R-charge two sector of the theory such that the generic F-terms force, in vacuum, the $u_{\alpha}$ to be set equal to these non-polynomial generators. For example, if $\phi$ denotes the fields associated to the original ambient space coordinates, F-flatness conditions of the schematic form $M(\phi) u_{\alpha} =f(\phi)$ force the new fields to take the value of a rational combination of those ambient space coordinates in vacuum. When this solution for the $u_{\alpha}$ is substituted into the remaining F-terms, the full non-polynomially deformed Calabi-Yau threefold defining relations are recovered. Because the rational nature of the non-polynomial deformations only appears in vacuum in this construction, the singularities that could appear elsewhere in field space (schematically at the locus $M(\phi)=0$) do not arise. Indeed, as can be seen from the form of the F-terms we have sketched, the superpotential of the GLSM is entirely polynomial in the fields.

In constructing non-polynomial deformations of complete intersection Calabi-Yau manifolds in this way, some of the original polynomial defining relations arise in a somewhat non-trivial manner: as non-trivial syzygies between the F-terms outlined above (when viewed as functions of the $u_{\alpha}$). As a result, the vacuum space of the GLSM including the non-polynomial deformations is in fact another complete intersection, albeit in some larger ambient variety.

It should be noted that there are similarities of this construction to the PAX and PAXY constructions of  \cite{Jockers:2012zr,Jockers:2012dk}. That work gives GLSM realizations of Pfaffian and determinantal varieties.  There also, the varieties themselves are not ordinarily described as complete intersections but, by going to a larger space, one is nevertheless able to give a GLSM realization.

\vspace{0.1cm}

While this work was in the final stages of preparation, \cite{Kanazawa:2026wqh,Kanazawa:2026zzr} appeared, which use a similar mathematical construction in a study of phases and mirror symmetry of gCICYs.

\vspace{0.1cm}

The rest of this paper is organized as follows. Section~\ref{npsec} begins with a review of the description of non-polynomial deformations of complete intersection Calabi-Yau threefolds. A description is then presented of finite non-polynomial deformations of Calabi-Yau manifolds in a GLSM context, illustrated with an explicit example. Section~\ref{sect:gcicy} presents the description of gCICYs in a GLSM context, again illustrated with an explicit example. Finally, in Section \ref{conc} we summarize and conclude. An illustration of our procedure for a class of examples is given in the Appendix.

\section{Non-polynomial deformations} \label{npsec}

\subsection{Generalities} \label{nonpolsec}

There are two types of non-polynomial deformation that can occur in a Calabi-Yau manifold $X$ described as a complete intersection in an ambient space $A$. Consider the  sequence
\begin{eqnarray}
    0 \to TX \to TA|_X \to {\cal N}_X \to 0 .
\end{eqnarray}
The long exact sequence in cohomology associated to this short exact sequence leads to the following extension describing how the complex structure moduli of the Calabi-Yau manifold are associated to kernel and cokernel pieces.
\begin{equation}  \label{h1tx}
    0 \to
    \text{coker} \!\left( H^0(X,TA|_X) \to H^0(X,{\cal N}_X)\right) 
    \to H^1(X,TX) \to
    \ker\! \left(H^1(X,TA|_X) \to H^1(X,{\cal N}_X) \right)
    \to 0.
\end{equation}
The right (ker) term describes complex structure moduli on $X$ that arise from complex structure moduli on $A$. That is, if the complex structure tensor ${\cal J}$ is changed on $A$, and then ${\cal J}$ is pulled back under the embedding map to $X$, this will result in a change in the complex structure on $X$. This type of non-polynomial deformation has been considered in a GLSM context in \cite{Adams:2023imc}. The left (coker) term in (\ref{h1tx}) lives inside the cohomology $H^0(X,{\cal N}_X)$. This cohomology is related to ambient space quantities by the Koszul complex
\begin{eqnarray}  \label{eq:koszul-res}
    0 \to \wedge^{K}{\cal N}^{\vee} \otimes {\cal N} \to \wedge^{K-1}{\cal N}^{\vee} \otimes {\cal N} \to \ldots \to {\cal N} \to {\cal N}_X \to 0 .
\end{eqnarray}
In this exact sequence of sheaves on $A$, $K$ is the codimension of $X$ in $A$ and ${\cal N}$ is the sum of line bundles on the ambient space that the defining equations of $X$ are sections of (the sum that restricts to ${\cal N}_X$ on $X$). By decomposing this sequence into short exact pieces and chasing the associated long exact sequences in cohomology, one can obtain 
\begin{equation} \label{nxcoh}
    0 \to {\rm coker}\! \left( H^0(A,{\cal I}_X \otimes {\cal N}) \to H^0(A, {\cal N}) \right) 
    \to H^0(X, {\cal N}_X)
    \to
    {\rm ker}\!\left( H^1(A, {\cal I}_X \otimes {\cal N}) \to H^1(A, {\cal N}) \right)
    \to 0.
\end{equation}
Here ${\cal I}_X$ is the ideal sheaf associated to $X$. When this description of $H^0(X,{\cal N}_X)$ is substituted into (\ref{h1tx}), the left (coker) term in (\ref{nxcoh}) corresponds to the polynomial complex structure deformations. That is, it details those complex structure moduli which are associated to elements of $H^0(A,{\cal N})$: polynomials in ambient space coordinates. The right (ker) term corresponds to the second of the two types of non-polynomial complex structure deformations mentioned above. It is these complex structure degrees of freedom that we will incorporate into a GLSM description for the first time in this paper. These non-polynomial complex structure deformations admit a description in terms of certain rational functions which are the same in spirit as the description of the gCICY Calabi-Yau manifolds \cite{Anderson:2015iia}.

Next we will review how this class of non-polynomial complex structure deformations can arise and be described in a concrete example. This same example will then be used to demonstrate the procedure for constructing a GLSM encompassing non-polynomial deformations. An outline of the more general procedure (for a class of examples) is given in Appendix \ref{illustration_procedure}.

\subsection{Geometry of an example} \label{nonpolyintro}

Consider the following CICY threefold.
\begin{eqnarray}\label{cicyx2}
    X_1 = \left[\begin{array}{c|cc} \mathbb{P}^1 & 0 & 2 \\ \mathbb{P}^1 & 1 & 1 \\\mathbb{P}^3&3&1\end{array} \right] .
\end{eqnarray}
Although the ambient space is rigid, this algebraic variety has non-polynomial deformations of the second type discussed in Section \ref{nonpolsec}.  In this example, the non-polynomial deformations are all associated to the first defining equation. The ${\cal O}_{X_1}(0,1,3)$ term in the normal bundle has 49 global sections, but only 39 of these are polynomial in the homogeneous coordinates of the ambient space. The remaining 10 global sections can be described (in a gCICY style construction) as follows.

First, to set notation, describe the complete intersection above as the vanishing locus
\begin{equation}
    P_{(2,1,1)} \: = \: 0 \: = \: P_{(0,1,3)},
\end{equation}
where $P_{(2,1,1)}$ has multi-degree $(2,1,1)$, and $P_{(0,1,3)}$ has multi-degree $(0,1,3)$.
Write out $P_{(2,1,1)}$ as
\begin{eqnarray}  \label{p211def}
    P_{(2,1,1)}=f_1 x_0^2+f_2 x_0 x_1 + f_3 x_1^2.
\end{eqnarray}
Here, the $f_i$ are general polynomials of multi-degree $(0,1,1)$ and $x_0$ and $x_1$ are the homogeneous coordinates of the first $\mathbb{P}^1$ factor in the ambient space. From (\ref{p211def}) we see that, on $X_1$, if $x_0=0$ then $f_3=0$ and similarly $x_1=0 \Rightarrow f_1 =0$. Therefore, we can define the following two rational functions which have no poles on $X_1$:
\begin{eqnarray}
    \frac{f_1}{x_1}\;\;\;,\;\;\; \frac{f_3}{x_0} .
\end{eqnarray}
These are therefore well defined functions of multi-degree $(-1,1,1)$ on $X_1$.
Multiplying these by polynomials of multi-degree $(1,0,2)$ results in well-defined rational functions of multi-degree $(0,1,3)$, which are therefore examples of non-polynomial deformations of $P_{(0,1,3)}$ and so the CICY. Furthermore, functions of this form are sufficient to recover all of the missing global sections of ${\cal O}_{X_1}(0,1,3)$ and so the general degree $(0,1,3)$ defining relation. Therefore, that defining relation can be written as the following deformation of the purely polynomial defining relation $p_1$:
\begin{eqnarray} \label{full031}
    P_{(0,1,3)}= p_1 + g_0 \frac{f_3}{x_0} + g_1\frac{f_1}{x_1} .
\end{eqnarray}

The gCICYs of \cite{Anderson:2015iia} can simply be thought of as CICYs where, for one or more of the defining relations, the analogue of $p_1$ in (\ref{full031}) is missing and all of the complex structure of those defining relations is non-polynomial. We will return to this subject in a later section.

A GLSM description of the non-polynomial deformations of the type described in this section may seem problematic.  After all, if there were contributions to the GLSM superpotential proportional to the rational form found in (\ref{full031}), the superpotential would be finite on the Calabi-Yau manifold but would have poles elsewhere in field space. We now turn to describing how this issue can be avoided and perfectly well defined GLSMs can be written down that encompass the non-polynomial deformations in (\ref{full031}).

\subsection{GLSM definition and vacuum space analysis}

We wish to construct a GLSM whose vacuum space in one phase is precisely the geometry described in the previous subsection, including its non-polynomial deformations. As mentioned at the end of that subsection, we cannot simply include a superpotential term proportional to (\ref{full031}) as that would result in singularities at certain loci in field space. 

Our approach will be schematically as follows. We will introduce new fields $u_0$ and $u_1$ that, as with all of the other fields, will only appear polynomially in the  GLSM superpotential. The charges of the GLSM fields will be chosen such that, for a generic superpotential, the F-terms set $u_0=f_3/x_0$ and $u_1=f_1/x_1$ {\it in vacuum}. One of the other F-terms that appears will take the form,
\begin{eqnarray}
  p_1+g_0 u_0 +g_1 u_1 =0  \;.
\end{eqnarray}
Thus, when the in-vacuum values for $u_0$ and $u_1$ are substituted into this F-term, the full non-polynomially deformed defining relation (\ref{full031}) is recovered. In this manner, the theory will be able to encompass non-polynomial complex structure deformations in a theory with a purely polynomial superpotential.

Note the way in which this construction avoids singularities in the GLSM field space. The rational functions that appear in describing non-polynomial complex structure deformations are smooth on the Calabi-Yau manifold, but have poles elsewhere in the ambient space. By having these functions only arise in the GLSM as a consequence of the vacuum equations, we ensure that they never appear in a context where their pole structure could come into play. Thus we end up with a perfectly well controlled GLSM.

While the above is schematically the mechanism we will utilize, there are many details which must be addressed to ensure that the GLSM exhibits this behavior at a generic point in its moduli space and not just for some tuned superpotential. For example, the defining relation $P_{(2,1,1)}$ which allows the definition of rational functions that are finite in vacuum is not associated to a superpotential term that reproduces (\ref{p211def}) directly. Rather, it is reproduced as a specific combination of the F-terms that appear in the field basis given below. Once all of these details are taken into account, we end up with a perfectly well defined GLSM, encompassing the non-polynomial deformations, where the Calabi-Yau vacuum space is described once again as a complete intersection, albeit in a larger ambient space.

\vspace{0.2cm}

Let us see all of this in detail. Consider a GLSM with gauge group $U(1)^3$ and field content as in Table \ref{fc1}.
\begin{table}[htbp]
\centering
\begin{tabular}{lccc}
\toprule
Field & $U(1)_1$ & $U(1)_2$ & $U(1)_3$ \\
\midrule
$x_0, x_1$ & $1$ & $0$ & $0$ \\
$y_0, y_1$ & $0$ & $1$ & $0$ \\
$z_0,\dots,z_3$ & $0$ & $0$ & $1$ \\
$u_0, u_1$ & $-1$ & $1$ & $1$ \\
$\Lambda_1, \Lambda_2, \Lambda_3$ & $0$ & $-1$ & $-1$ \\
$P$ & $0$ & $-1$ & $-3$ \\
\bottomrule
\end{tabular}
\caption{GLSM field content}
\label{fc1}
\end{table}
Here the fields $\Lambda_i$ and $P$ have R-charge $2$ and all other degrees of freedom have vanishing R-charge. Up to field redefinitions, the most general superpotential that can be written down in such a theory is as follows:
\begin{eqnarray} \label{Weg1}
    W =&& \Lambda_1\big(x_0 u_0 - f_3(y,z)\big) + \Lambda_2\big(x_1 u_1 - f_1(y,z)\big) + \Lambda_3\big(x_1 u_0 + x_0 u_1 + f_2(y,z)\big) \\ \nonumber
&&+ P\big(p_1(y,z) + g_0(x,z)\,u_0 + g_1(x,z)\,u_1\big) .
\end{eqnarray}

The ideas described above behind the GLSM can already be seen from this expression for $W$. The naive solution to the F-terms for $\Lambda_1$ and $\Lambda_2$ is given by
\begin{eqnarray}
u_0 =\frac{f_3}{x_0} \;\;,\;\; u_1 = \frac{f_1}{x_1} .
\end{eqnarray}
If these fixed values for the $u$'s (given values for the other fields) are then substituted into the F-term associated to $P$ we find exactly our non-polynomially deformed defining relation (\ref{full031}). 
The above sketch is morally correct, however, a careful vacuum analysis is required to ensure that everything is working as desired.

Let us take all three FI parameters in the theory to be positive. The first of these, associated to $U(1)_1$ from Table \ref{fc1}, immediately implies that $x_0$ and $x_1$ can not simultaneously vanish. Writing the F-terms associated to the $\Lambda_i$ fields in matrix form,
\begin{eqnarray} \label{ueq}
 \left(  \begin{array}{cc} x_0 & 0 \\ x_1 & x_0 \\ 0 & x_1 \end{array} \right) \left( \begin{array}{c} u_0 \\ u_1 \end{array}\right) = \left(\begin{array}{c} f_3 \\ -f_2 \\ f_1 \end{array} \right) \;,
\end{eqnarray}
then leads to the conclusion that the matrix of $x$'s appearing always has maximal rank and therefore $u_0$ and $u_1$ are uniquely determined everywhere in vacuum. If the $y$'s could all simultaneously vanish, the $f_i$ would all simultaneously vanish and from (\ref{ueq}) so would the $u$'s. In such a situation the $U(1)_2$ D-term could then not be satisfied, leading to the conclusion that a combination of our choice of FI parameters and the F-terms leads to the $y$'s not all simultaneously vanishing in vacuum. A similar argument can be made for the $z$'s. Thus our choice of FI parameters ensure that the $x$'s, $y$'s and $z$'s all form the homogeneous coordinates of projective spaces when we restrict to the F-flat vacuum.

To determine the values to which equation~(\ref{ueq}) fixes the $u$'s,
we consider three possible cases.
\begin{itemize}
    \item Case 1: $x_0\neq0$ and $x_1 \neq0$. In this case clearly we have 
    \begin{eqnarray}
        u_0=f_3/x_0 \;\;\; \text{and} \;\;\; u_1=f_1/x_1\;\;,
    \end{eqnarray}
    from solving the F-terms for $\Lambda_1$ and $\Lambda_2$ respectively. The F-term for $\Lambda_3$ then implies that
    \begin{eqnarray}
        x_1f_3/x_0+ x_0f_1/x_1 + f_2 =0 \Leftrightarrow P_{(2,1,1)}=0\;.
    \end{eqnarray}
    The F-term for $P$ then correctly reproduces the degree (0,1,3) defining relation, including the gCICY style description of the non-polynomial deformations
    \begin{eqnarray} \label{npdef1}
        p_1+ g_0 \frac{f_3}{x_0} + g_1 \frac{f_1}{x_1} =0 .
    \end{eqnarray}
    \item Case 2: $x_0=0$ and $x_1 \neq0$. In this case we find that 
    \begin{eqnarray}
        u_1=f_1/x_1 \;\;\; \text{and} \;\;\; f_3=0
    \end{eqnarray} in vacuum, from the F-terms for $\Lambda_1$ and $\Lambda_2$. The F-term for $\Lambda_3$ now implies that $u_0=-f_2/x_1$. Note that if we consider $P_{(2,1,1)}=0$ on the locus where $x_0=0$ we do indeed find $f_3=0$. Thus we again have two uniquely specified $u$'s in vacuum and the correct degree $(2,1,1)$ defining relation. The F-term for $P$ then becomes
    \begin{eqnarray}\label{npdef2}
        p_1-g_0 f_2/x_1 + g_1 f_1/x_1\;.
    \end{eqnarray}
    Note that this is the correct form for the vacuum geometry we desire. From the degree $(2,1,1)$ defining relation we can write, on the patch where $x_1\neq0$, 
    \begin{eqnarray}
        \frac{f_3}{x_0}=-\frac{f_1 x_0^2+ f_2 x_0 x_1}{x_0 x_1^2}=-\frac{f_1 x_0+ f_2  x_1}{ x_1^2}\;.
    \end{eqnarray}
    In the limit where $x_0\to 0$ we see that we obtain $f_3/x_0 \to -f_2/x_1$, explaining the seeming discrepancy between (\ref{npdef1}) and (\ref{npdef2}).
    \item Case 3: $x_0\neq0$ and $x_1=0$. This case exactly mirrors case 2 above in its analysis.
\end{itemize}

We thus see that the D-terms and the F-terms associated to the R-charge $2$ fields are enough to ensure that the vacuum geometry of our GLSM is exactly the CICY we desired including the non-polynomial deformations, assuming the vacuum expectation values of those R-charge $2$ fields are themselves forced to vanish. It is trivial to show that this is the case, with the vanishing being enforced by the remaining $10$ F-terms.

\subsection{The limit without non-polynomial deformations}
 \label{polylim}

Consider tuning the coefficients appearing in the polynomials $g_0$ and $g_1$ in (\ref{Weg1}) to zero, and then integrating out any massive degrees of freedom that appear. Under such a procedure, we would expect to recover the usual GLSM for the geometry (\ref{cicyx2}) covering the subspace of complex structure moduli space where all non-polynomial deformations are set to zero. This is indeed the case, as we now describe.

If we set $g_0=g_1=0$, the F-terms for the $u$ fields are as follows:
\begin{eqnarray} \label{lambdaeq}
    \left( \Lambda_1,\Lambda_2,\Lambda_3 \right)\cdot \left(  \begin{array}{cc} x_0 & 0\\ 0 & x_1 \\ x_1 & x_0  \end{array} \right) =0 .
\end{eqnarray}
The three by two matrix appearing in that expression has maximal rank everywhere in the geometric phase, and thus these F-terms fix the value of two of the $\Lambda$'s (they are massive). Integrating out those fields that have their values fixed in terms of the other degrees of freedom therefore removes two $\Lambda$'s and the $u$'s.

The combination of $\Lambda$'s which remains is that in the kernel of the transpose of the matrix appearing in (\ref{lambdaeq}). This can be written as
\begin{eqnarray} \label{lambdadef}
    \left( \Lambda_1,\Lambda_2,\Lambda_3 \right)= \lambda \left( -x_1^2,-x_0^2,x_0 x_1 \right) .
\end{eqnarray}

The degree of freedom $\lambda$ then has charge $(-2,-1,-1)$. Substituting (\ref{lambdadef}) back into (\ref{Weg1}), we are left with
\begin{eqnarray}
W= P p_1(y,z) + \lambda P_{(2,1,1)}\;.
\end{eqnarray}
That is, we arrive back at the standard GLSM for this geometry, with precisely the usual field content and superpotential. 

In short, for large values of the $U(1)_1$ FI parameter, which causes the $x$'s to acquire large expectation values, the $u$'s and two of the $\Lambda$'s become massive and integrating them out gives back the standard GLSM not including non-polynomial deformations. 

\subsection{Charge balance in the enlarged GLSM} \label{anom}

It may not be obvious how a procedure such as the one we have described always leads to a well defined GLSM. In particular, in comparison to a standard GLSM which only describes the polynomial deformations of some complete intersection, we have removed some fields and added others. It may, therefore, not be clear that the $U(1)$ charges still add up to zero correctly in each case. We will now show that they do.

The mechanism we have described proceeds by replacing some superpotential terms describing polynomial defining equations, such as the one of degree $(2,1,1)$ in the preceding example, by several terms which give rise to F-terms in the following form.

\begin{eqnarray} \label{genf}
F_i=\sum_{\alpha=1}^r M_{i \alpha} u_{\alpha} - f_i =0
\end{eqnarray}
Here, the number $r$ of $u_{\alpha}$'s appearing would be determined by the number of generating fractions of the non-polynomial deformations. The number, $r+t$, of new F-terms (the index range of $i$) is determined by $r$ and the number of original defining relations we are replacing, $t$. This is because the dimension of the vacuum space should not change and we therefore need an extra $r$ F-terms to fix the additional $r$ coordinates $u$.

The consistency of our GLSMs relies on two features of the construction. The first is that $\rank M=r$ at every point on the vacuum locus so that the $u$ fields are uniquely specified in terms of the other coordinates there. The second is that there exists a $t \times (r+t)$ matrix ${\cal K}$ such that ${\cal K} M=0$ so that we can form combinations  ${\cal K} F=-{\cal K} f$ of the F-terms which are those that reproduce the original polynomial F-terms we have replaced (see the discussion in Section \ref{polylim}). For the argument below we will also need that $\rank M=r$ and $\rank {\cal K}=t$ away from a locus of codimension at least two in the ambient space. This holds for generic coefficient choices in the superpotential. The degeneracy locus of $M$ is of codimension $t+1 \geq 2$ in such a case, and ${\cal K}$ can be chosen to degenerate only where $M$ does.

Let us call the charges of the $u_{\alpha}$ fields $q_{\alpha}$, those of the new R-charge 2 fields $-D_i$ and those of the R-charge 2 fields that our construction replaces $-d_A$ (where $A=1\ldots t$). Note that all of these charges are in fact vectors when multiple $U(1)$s are present, but we will suppress this structure in what follows as the argument works for each $U(1)$ independently. With this nomenclature, the above discussion shows we can write the following sequence of vector bundles over the ambient space of the original Calabi-Yau construction whose non-polynomial deformations we are describing.
\begin{eqnarray}
0\to \bigoplus_{\alpha=1}^r {\cal O}(q_{\alpha}) \stackrel{M}{\longrightarrow} \bigoplus_{i=1}^{r+t} {\cal O}(D_i) \stackrel{{\cal K}}{\longrightarrow} \bigoplus_{A=1}^t {\cal O}(d_A) \to 0
\end{eqnarray}
Note that this sequence is exact in the middle, away from a locus of codimension at least two, because $\operatorname{im}(M) \subseteq \ker({\cal K})$ are subbundles of equal rank there. Because this sequence is short exact, away from codimension two, the alternating sum of the first Chern classes vanishes. This follows from the fact that the sheaves that would be needed to complete it to a sequence that is exact over every point of the ambient space would have vanishing first Chern class due to the codimension of their support. In the case at hand this says that,
\begin{eqnarray}
\sum_{\alpha} q_{\alpha} -\sum_i D_i = -\sum_A d_A\;.
\end{eqnarray}
In words, the total charge of the fields that have been added is equal to the total charge of the fields that have been removed. Thus, if the original ``polynomial deformation only'' GLSM had field content such that the charges summed to zero, so does the construction we have presented.

\section{gCICYs}  \label{sect:gcicy}
 
Exactly the same trick that we used in the previous section to construct a GLSM encompassing non-polynomial deformations of a CICY can be used to construct well behaved GLSMs associated to gCICYs. The concept of gCICYs, first introduced in \cite{Anderson:2015iia}, is one in which {\it all} of the complex structure degrees of freedom associated to a given subset of defining equations are non-polynomial in nature. In this sense, these are an extreme form of the ordinary CICYs which admit non-polynomial deformations.

\subsection{Geometry of an example}  \label{secgcicyintro}

The easiest way to illustrate the gCICY concept is with an example. Consider a family of smooth Calabi-Yau threefolds defined by the following configuration matrix.
\begin{eqnarray} \label{gcicyconf}
    X_2 = \left[ \begin{array}{c||cc|c} \mathbb{P}^1 & 0&1 & 1 \\ \mathbb{P}^1&1&2&-1 \\ \mathbb{P}^4 & 1&1&3 \end{array} \right] .
\end{eqnarray}
The first column of (\ref{gcicyconf}) describes the ambient space $\mathbb{P}^1 \times \mathbb{P}^1 \times \mathbb{P}^4$. The next two columns of the configuration matrix define the multi-degree of defining polynomials of a variety, ${\cal M}$, just as would be the case for an ordinary CICY. However, the variety ${\cal M}$ obtained by applying just the $(0,1,1)$ and $(1,2,1)$ equations is not a Calabi-Yau manifold. The final column defines the multi-degree of a defining relation as well, but here negative entries are allowed. The idea is that this defining relation is a section of the line bundle ${\cal O}_{\cal M}(1,-1,3)$. While this line bundle has no global sections on $\mathbb{P}^1 \times \mathbb{P}^1 \times \mathbb{P}^4$ it does have them on ${\cal M}$. This defining relation therefore cannot be described as a polynomial in the ambient space coordinates, just like the non-polynomial complex structure deformations discussed in the previous section. The difference in the gCICY case is that there is {\it no} polynomial defining relation of this degree, not merely some deformations that can not be described as ambient space polynomials. The configuration matrix \eqref{gcicyconf} defines a smooth CY threefold with Hodge numbers $(h^{1,1},h^{2,1})=(3,48)$.

The defining relations of gCICYs can be thought of, more naively, as rational functions of the given mixed-sign multi-degree \cite{Anderson:2015iia}. The rational defining relation in the current case, for example, can be constructed explicitly as follows. Write the degree $(0,1,1)$ and $(1,2,1)$ defining relations as 
\begin{eqnarray} \label{one}
    P_{(0,1,1)} &=& l_0(z) y_0+l_1(z)y_1 , \\ \label{two}
    P_{(1,2,1)} &=& L_0(x,z) y_0^2+ L_1(x,z) y_0 y_1 + L_2(x,z) y_1^2 .
\end{eqnarray}
In these expressions $x,y$ and $z$ are the homogeneous coordinates on the first, second and third ambient space $\mathbb{P}^n$ factors respectively. From (\ref{one}), we see that on ${\cal M}$, if $y_0=0$ then $l_1=0$ and similarly $y_1=0 \Rightarrow l_0 =0$. Therefore, we can define the following two rational functions, which are related by $P_{(0,1,1)}=0$, and which have no poles on ${\cal M}$:
\begin{eqnarray}
\frac{l_0}{y_1} = - \frac{l_1}{y_0} .
\end{eqnarray}
Performing a similar analysis on $P_{(1,2,1)}$ we find the following independent rational functions
\begin{eqnarray}
    \frac{L_0}{y_1}\;\;\; \text{and} \;\;\; \frac{L_2}{y_0}.
\end{eqnarray}
Thus we have three independent rational functions which do not have poles on ${\cal M}$, with degrees $(0,-1,1)$, $(1,-1,1)$ and $(1,-1,1)$ respectively. By multiplying each of these by a general polynomial of the correct degree and summing the results, we can find the rational defining relation of degree $(1,-1,3)$ that appears in (\ref{gcicyconf}) which, in this case, can be proven to describe a smooth Calabi-Yau threefold \cite{Anderson:2015iia}
\begin{eqnarray} \label{ratdef}
    P_{(1,-1,3)}=\frac{l_0}{y_1} g_0^{(1,0,2)} + \frac{L_0}{y_1} g_1^{(0,0,2)} +  \frac{L_2}{y_0} g_2^{(0,0,2)}.
\end{eqnarray}
Here, the $g_i$'s are polynomials with the multi-degree indicated.

Note that, as for the non-polynomial deformations of CICYs described earlier in this paper, the crucial feature of the above construction is that, while the rational function (\ref{ratdef}) does have poles, they miss the manifold ${\cal M}$ and thus the Calabi-Yau manifold embedded inside it. Once again, one could imagine that this might cause problems in a GLSM construction where the ambient space coordinates become fields in the theory, and thus one might expect singularities in the wider field space. As before, we resolve this issue by describing the gCICY as an ordinary complete intersection in a larger ambient space, following a pattern essentially identical to that which we used earlier to describe non-polynomial deformations.

\subsection{GLSM definition and vacuum space analysis} \label{gcicysec}

Let us specify the GLSM that describes the gCICY example given in (\ref{gcicyconf}), and reproduced here:

\begin{eqnarray} \label{gcicyconf2}
    X_2 = \left[ \begin{array}{c||cc|c} \mathbb{P}^1 & 0&1 & 1 \\ \mathbb{P}^1&1&2&-1 \\ \mathbb{P}^4 & 1&1&3 \end{array} \right].
\end{eqnarray}

Consider a GLSM with gauge group $U(1)^3$ and field content as in Table \ref{fc2}. 
\begin{table}[htbp] \centering \begin{tabular}{lccc} \toprule Field & $U(1)_1$ & $U(1)_2$ & $U(1)_3$  \\ \midrule $x_0, x_1$ & $1$ & $0$ & $0$  \\ $y_0, y_1$ & $0$ & $1$ & $0$  \\ $z_0,\dots,z_4$ & $0$ & $0$ & $1$  \\ $w$ & $0$ & $-1$ & $1$ \\ $v_0, v_1$ & $1$ & $-1$ & $1$  \\ $\Lambda'_1, \Lambda'_2$ & $0$ & $0$ & $-1$  \\ $\Lambda''_1, \Lambda''_2, \Lambda''_3$ & $-1$ & $0$ & $-1$  \\ $P$ & $-1$ & $1$ & $-3$ \\ \bottomrule \end{tabular} \caption{GLSM field content} \label{fc2} \end{table}
Here the fields $\Lambda_i'$, $\Lambda_i''$ and $P$ have R-charge 2 and all other degrees of freedom have vanishing R-charge. Up to field redefinitions, the most general superpotential that can be written down in such a theory is 
\begin{eqnarray} \label{sup2}
W \;&=&\;\
\Lambda'_1\big(w\,y_0 - l_1\big) + \Lambda'_2\big(w\,y_1 + l_0\big)
+ \Lambda''_1\big(v_0\,y_0 - L_2\big) + \Lambda''_2\big(v_1\,y_1 - L_0\big)
 \\ \nonumber &&+ \Lambda''_3\big(v_0\,y_1 + v_1\,y_0 + L_1\big) 
+ P\big(-g_0(x,z)\,w + g_1(z)\,v_1+ g_2(z)\,v_0 \big).
\label{eq:WII}
\end{eqnarray}
As in the preceding subsection, the idea behind this GLSM is to introduce new variables, here $w$, $v_0$ and $v_1$ that are set equal to the rational quantities of interest by the F-flatness conditions.

\vspace{0.2cm}

Let us take all three FI parameters to be positive. Now, suppose that $y_0=y_1=0$. In such a case, the sum of the $U(1)_2$ and $U(1)_3$ D-terms would then imply that not all the $z_i$ can simultaneously vanish and the $U(1)_2$ D-term on its own would imply that $P$ does not vanish. Let us assume in what follows that we make generic choices for the polynomials appearing in (\ref{sup2}). The F-terms associated to the $\Lambda'$'s and $v_i$'s would then give us four independent functions of $z$ which would have to vanish - which fixes the possible $z$ values to a finite set of points. The F-terms associated to the $\Lambda''$'s give three linear constraints in $x$ at each such point so we also then have that $x_0=x_1=0$. Finally, the F-terms associated to the $z$ fields are then constraining enough to force $v_0$ and $v_1$ to vanish. However, if the $v$'s and the $x$'s vanish then the D-term associated to $U(1)_1$ can not be satisfied, and so we are led to the conclusion that the original assumption, $y_0=y_1=0$, is not possible.

From the analysis of the preceding paragraph we learn that a combination of the D- and F-terms imply that the $y$'s can not all simultaneously vanish. The $\Lambda'$ and $\Lambda''$ F-term constraints can be written as
\begin{eqnarray} \label{secondf}
    \left( \begin{array}{c} y_0 \\ y_1 \end{array}\right)w=\left( \begin{array}{c} l_1\\ -l_0 \end{array} \right) \;\;\; \text{and} \;\;\; \left( \begin{array}{cc}  y_0&0\\0&y_1\\ y_1&y_0\end{array} \right)\left(\begin{array}{c} v_0\\ v_1 \end{array} \right)=\left(\begin{array}{c} L_2\\ L_0 \\ -L_1\end{array} \right),
\end{eqnarray}
respectively. Since the $y_i$'s can't all vanish, these equations uniquely determine $w,v_0$ and $v_1$. 

Imagine a case where all of the $x$'s simultaneously vanish. In such a situation, the $L_i=0$ and (\ref{secondf}) then implies that $v_0=v_1=0$. If both the $x$'s and the $v$'s vanish then the $U(1)_1$ D-term can not be satisfied at positive FI parameter and so we conclude the $x$'s can't all simultaneously vanish.

If all of the $z$'s were to simultaneously vanish, then $l_0=l_1=L_i=0$. In such a situation (\ref{secondf}) would mean that $w=v_0=v_1=0$ as well, which would mean that the $U(1)_3$ D-term could not be solved.

The conclusion of the last several paragraphs is that taking all three FI parameters to be positive implies that the origin of field space for the $x$'s, $y$'s and $z$'s each individually is removed from the vacuum space of the theory. The F-terms (\ref{secondf}) then imply that the values of $w$, $v_0$ and $v_1$ are fixed given these coordinates. Note that the integrability conditions on the overconstrained equations in (\ref{secondf}) imply the following two equations:
\begin{eqnarray}
    l_0 y_0 + l_1 y_1=0 \;\; \; \text{and} \;\;\; L_0 y_0^2+ L_1 y_0 y_1 +L_2 y_1^2 =0 .
\end{eqnarray}
These are nothing but the two polynomial relations defining $X_2$.

To determine the values to which equation~(\ref{secondf}) fixes the $v$'s and $w$'s,
we consider
three possible cases.
\begin{itemize}
    \item Case 1: $y_0 \neq0$ and $y_1 \neq 0$. In this case we clearly have
\begin{eqnarray}
    w=-\frac{l_0}{y_1} = \frac{l_1}{y_0} \;\; ,\;\; v_0= \frac{L_2}{y_0} \;\; \text{and} \;\; v_1=\frac{L_0}{y_1} .
\end{eqnarray}
The F-term for $P$ then correctly reproduces the mixed sign defining relation of the gCICY as presented in (\ref{ratdef})
\begin{eqnarray}
    g_0 \frac{l_0}{y_1}+ g_1\frac{L_0}{y_1} + g_2 \frac{L_2}{y_0}=0 .
\end{eqnarray}
\item Case 2: $y_0=0$ and $y_1\neq0$. The F-terms associated to $\Lambda_1'$ and $\Lambda_1''$ imply that $l_1=L_2=0$ in this case, which is exactly the form the two polynomial defining relations of $X_2$ take when $y_0=0$. The F-terms associated to $\Lambda_2'$, $\Lambda_2''$ and $\Lambda_3''$ then imply 
\begin{eqnarray}
    w= -\frac{l_0}{y_1} \;\;,\;\; v_1= \frac{L_0}{y_1} \;\; \text{and} \;\; v_0 = -\frac{L_1}{y_1} .
\end{eqnarray}
Thus we again have uniquely specified $v$'s and $w$ in vacuum and the F-term for P then becomes
\begin{eqnarray} \label{eg2case2}
g_0 \frac{l_0}{y_1}  + g_1 \frac{L_0}{y_1}-g_2 \frac{L_1}{y_1}=0 \;.
\end{eqnarray}
Note that when $y_0=0$ we have
\begin{eqnarray}
    \frac{L_2}{y_0} = -\frac{L_0 y_0^2 + L_1 y_0 y_1}{y_0 y_1^2} \to -\frac{ L_1}{ y_1} 
\end{eqnarray}
so that (\ref{eg2case2}) is indeed the correct limiting form of the gCICY defining relation with negative entry.
\item Case 3: $y_0 \neq 0$ and $y_1=0$. This case exactly mirrors case 2 above in its analysis.
\end{itemize}

We thus see that the D-terms and the F-terms associated to the R-charge 2 fields are enough to ensure that the vacuum geometry of our GLSM is exactly the gCICY we desired, assuming that the vacuum expectation values of those R-charge 2 fields are themselves forced to vanish. It is trivial to show that this is the case, with the vanishing
being enforced by the remaining 12 F-terms, for sufficiently generic choices of the polynomials appearing in (\ref{sup2}).

\vspace{0.2cm}

In short, precisely the same construction that allowed us to accommodate non-polynomial complex structure deformations in a GLSM context can also describe gCICYs.

\vspace{0.2cm}

With the toric description of \eqref{gcicyconf} associated to the charge matrix given in Table \ref{fc2} in hand, another non-trivial check of our procedure can be undertaken -- namely the computation of the Hodge numbers, triple intersection numbers and K\"ahler/Mori cone data of $X_2$, now computed directly as a toric complete intersection manifold rather than a gCICY. This check was undertaken using standard tools (e.g.~SageMath, PALP) and found to agree exactly with the same numbers obtained from the gCICY description of $X_2$, using the techniques of \cite{Anderson:2015iia}.

\section{Summary and conclusions} \label{conc}

In this paper we have described how to incorporate non-polynomial complex structure deformations,
of the type not inherited from the ambient space, into the GLSM description of complete intersection
Calabi-Yau manifolds. Our formalism works at the level of finite deformations, not just infinitesimal
perturbations. As a natural limit of this construction, in which an entire defining relation is non-polynomial, we have also shown how to construct well defined GLSMs describing gCICYs.

In our approach, the rational functions that describe non-polynomial complex structure deformations only appear upon solving the F-terms to find the vacuum space. This allows us to retain
an entirely polynomial superpotential, the most general one compatible with the charges of the
theory up to field redefinitions, which therefore has no poles anywhere in field space. We have also shown how setting the non-polynomial complex structure deformations to zero and integrating out massive states recovers the standard GLSM for these Calabi-Yau manifolds. 

There are many possible directions for future work following from these results. The phase
structure of gCICYs and of non-polynomially deformed CICYs can now be explored in their GLSM
realizations (see \cite{Kanazawa:2026wqh,Kanazawa:2026zzr} for related work). We expect the generalization of our construction to more general toric ambient spaces
to be straightforward. Finally, similar techniques can be used to study non-polynomial deformations of two-dimensional (0,2) supersymmetric theories \cite{Distler:1993mk,Melnikov:2019tpl}. We
will pursue these, and other, directions in forthcoming work.

\section*{Acknowledgements}

E.S.~would like to thank I.~Melnikov for useful discussions. L.A.~and J.G.~would like to thank S.~J.~Lee and C.~Closset for initial discussions on this topic many years ago which planted the seed for the authors to return to the subject. 
L.A., J.G., and E.S.~were partially supported by NSF grant PHY-2609668. The authors employed Claude Code and ChatGPT/Codex in developing this work; however, we are responsible for all claims and for the writing.

\appendix

\section{Construction for a class of non-polynomial deformations}\label{illustration_procedure}

In this appendix we summarize, for a broad class of examples, the procedure illustrated in the preceding sections. For simplicity, we restrict attention to non-polynomial sections whose non-polynomial structure is associated to a single $\mathbb{P}^1$ factor of the ambient space (defined below in terms of contributions to cohomology arising in the Koszul sequence).

Consider a manifold defined as a complete intersection in a product of $F$ projective spaces, as discussed in Section~\ref{npsec}.
Specifically, let
\begin{equation}
    A \: = \: \prod_{s=1}^F {\mathbb P}^{n_s}, \: \: \: X \: = \: \{ P_1 = \cdots = P_K = 0\},
    \: \: \:
    P_k \in H^0(A, {\cal O}(\vec{d}_k)),
\end{equation}
with $\vec{d}_k=(d_k^1, \cdots, d_k^F)$. In the case that the complete intersection is a Calabi-Yau threefold, $X$, the familiar degree requirement is
\begin{equation}
    \sum_{k=1}^K d_k^s  \: = \: n_s+1. 
\end{equation}
 In the gCICY case we instead denote by ${\cal M}=\{P_1=\cdots=P_K=0\}$ the (non-Calabi-Yau) complete intersection, and $X={\cal M}\cap\{P_{\ve}=0\}$ with $P_{\ve}\in H^0({\cal M},\cO_{\cal M}(\ve))$ and $H^0(A,\cO(\ve))=0$. The Calabi-Yau condition then reads $\sum_k \vd_k+\ve=(n_1+1,\dots,n_F+1)$.
 
The deformations we are studying, in the language of the previous sections, originate as elements of $H^1(A, {\cal I}_X \otimes {\cal N})$.  We will use the Koszul resolution of ${\cal I}_X \otimes {\cal N}$ (see (\ref{eq:koszul-res})), which we write as
\begin{equation}
      0 \to \wedge^{K}{\cal N}^{\vee} \otimes {\cal N} \to \wedge^{K-1}{\cal N}^{\vee} \otimes {\cal N} \to \ldots \to {\cal N}^{\vee} \otimes {\cal N} \to {\cal I}_X \otimes {\cal N} \to 0.
\end{equation}
The cohomology group $H^{\bullet}(A, {\cal I}_X \otimes {\cal N})$ can be computed using the spectral sequence \cite[section 3.5]{gh}
\begin{equation}
    E_1^{-p,q} = H^q(A, \wedge^{p+1} {\cal N}^{\vee} \otimes {\cal N}) \: \Longrightarrow \: H^{q-p}(A, {\cal I}_X \otimes {\cal N} ),
\end{equation}
from which we see that terms that contribute to $H^1(A, {\cal I}_X \otimes {\cal N})$ are of the form
\begin{equation}
    H^p\left( A, \wedge^p {\cal N}^{\vee} \otimes {\cal N} \right)~~\text{for}~p \geq 1~.
\end{equation}
For simplicity, in this section we will only consider contributions from terms with $p=1$. It should be noted however that the description of other classes of non-polynomial sections has been studied in \cite{Anderson:2015iia} and we anticipate that the GLSM constructions we provide below can be naturally extended to this context as well.
For the $p=1$ case we consider here, using 
\begin{equation}
    \wedge^1 {\cal N}^{\vee} \otimes {\cal N} \: = \: \bigoplus_{j,k} {\cal O}(\vec{d}_j - \vec{d}_k ),
\end{equation}
and the Bott formula \cite[section 1.1]{oss}, $H^1\big(A,\cO(\vd_j-\vd_k)\big)$ can only be non-zero if there is a single $\mathbb{P}^1$ factor, in position $\sst$, such that
\begin{equation}
\begin{gathered}
h^1\big(A,\cO(\vd_j-\vd_k)\big)=\big(d_k^{\sst}-d_j^{\sst}-1\big)\prod_{s\neq\sst}h^0\big(\mathbb{P}^{n_s},\cO(d_j^s-d_k^s)\big)\,,\\
d_k^{\sst}-d_j^{\sst}\ge2\,,\qquad d_j^s-d_k^s\ge0\ \ (s\neq\sst)\,.
\end{gathered}
\label{eq:A-bott}
\end{equation}
Such a term corresponds to a deformation of $P_j$ by non-polynomial sections built from the \emph{source relation} $P_k$. Since these are $E_1$ terms, the expression in \eqref{eq:A-bott} only gives an upper bound on the number of first-order non-polynomial moduli (in the spectral sequence, differentials can reduce the number of actual moduli, or equivalently there can be non-trivial relations amongst the non-polynomial sections). For CICYs the entries in a $\mathbb{P}^1$ row sum to $2$. A contribution therefore requires $d_k^{\sst}=2$, with all other entries in that row vanishing. For gCICYs the analogous statement is that sections of $\cO_{\cal M}(\ve)$ that do not descend from $A$ arise, at this level of the spectral sequence, from $H^1\big(A,\cO(\ve-\vd_k)\big)$. In that case the negative entries of $\ve$ allow other values of $d_k^{\sst}$.

We will distinguish pairs $(\sst,k)$, where $\sst$ labels the ${\mathbb P}^1$ and $k$ is a column such that 
\begin{equation}
    h^1(A, {\cal O}(\vec{d}_j - \vec{d}_k) ) \neq 0
\end{equation}
for some $j$ (or with $\vec{d}_j$ replaced by $\vec{e}$ in the gCICY case). For such a pair $(\sst,k)$, write $x = (x_0,x_1)$ for the homogeneous coordinates on the
${\mathbb P}^1$.  Expand $P_k$ in $x$:
\begin{equation}
    P_k \: = \: \sum_{i=0}^{m} f_i \, x_0^{m-i} x_1^i,
\end{equation}
for $m = d_k^\sst$ and $f_i \in H^0(A, {\cal O}(d_k^1, \cdots, d_k^{\sst-1}, 0, d_k^{\sst+1}, \cdots, d_k^F))$.

Now, we are ready to describe a new presentation of $X$, in which the previously non-polynomial deformations become polynomial.
We replace the ambient space $A$ with a new ambient space $B$, 
a new toric variety constructed from $A$ in which the $u_i$ are added as new homogeneous (Cox) coordinates, of weights
\begin{equation} \label{eq:cox-wt-us}
   \vec{q}= \left( d_k^1, \cdots, d_k^{\sst-1}, -1, d_k^{\sst+1}, \cdots, d_k^F \right).
\end{equation}

We replace the original complete intersection
\begin{equation}
    \{ P_1 = P_2 = \cdots = P_K = 0 \} \subset A
\end{equation}
with a complete intersection of the following hypersurfaces in $B$:
\begin{equation} \label{eq:newci}
\begin{array}{c}
    P_1  =  0 = P_2 = \cdots P_{k-1} = P_{k+1} = \cdots = P_K,
    \\
    x_0 u_{i-1} - x_1 u_i \: = \: f_i, \: \: \: (i = 0, \cdots, m, \: \: \: u_{-1} = 0 = u_m).
    \end{array}
\end{equation}

Although the hypersurface $\{ P_k = 0 \}$ was omitted, it is nevertheless implicitly realized via a syzygy. That is,
\begin{eqnarray}
    P_k & = & \sum_{i=0}^m f_i x_0^{m-i} x_1^i,
    \\
    & = & \sum_{i=0}^m \left( x_0 u_{i-1} - x_1 u_i \right) x_0^{m-i} x_1^i,
    \\
    & = &
    (- x_1 u_0) x_0^m \: + \: ( x_0 u_0 - x_1 u_1) x_0^{m-1} x_1
    \: + \: (x_0 u_1 - x_1 u_2) x_0^{m-2} x_1^2 \: + \: \cdots.
\end{eqnarray}
However, this is a telescoping series, meaning neighboring terms cancel.  As a result, we see that
$P_k = 0$.  Conversely, so long as $x_0$ and $x_1$ do not simultaneously vanish, equations~(\ref{eq:newci}) uniquely determine the $u_i$ in terms of the $f_i$.  Explicitly,
\begin{equation}\label{udefs}
    u_a \: = \: - \frac{ \sum_{i \leq a} f_i x_0^{a-i} x_1^i }{ x_1^{a+1} }
    \: = \:
    \frac{ \sum_{i > a} f_i x_0^{m-i} x_1^{i-a-1} }{ x_0^{m-a} },
\end{equation}
where equality of the two expressions follows from $P_k=0$.

\vspace{0.2cm}

With these tools in hand, it is now possible to provide a description -- given the GLSM for the original complete intersection -- of the GLSM for the new presentation. We modify/augment the field content as follows:
\begin{itemize}
    \item Add new chiral superfields $u_a$, $a \in \{0, \cdots, m-1\}$, of charges~(\ref{eq:cox-wt-us}),
    \item Remove the field $p_k$, that multiplied $P_k$ in the superpotential,
    \item Add new fields $\Lambda_i$, $i \in \{0, \cdots, m\}$, of charge minus that of $f_i$,
    \item In the gCICY case, add a field $p_{\ve}$ of charge $-\ve$.
\end{itemize}
and the superpotential of the new GLSM becomes
\begin{eqnarray}
W & = &
p_1 P_1 + p_2 P_2 + \cdots + p_{k-1} P_{k-1} + p_{k+1} P_{k+1} + \cdots + p_K P_K
\\
& &
\: + \: \sum_{i=0}^m \Lambda_i \left( x_0 u_{i-1} - x_1 u_i - f_i \right) \Big[+\,p_{\ve}\,P_{\ve}\Big]\,.
\end{eqnarray}
(where the last term involving $P_{\ve}$ denotes an addition in the gCICY case). We note that the sum of the $U(1)$ charges of the $m$ $\{ u_a\}$ and the $m+1$ $\{ \Lambda_i \}$ matches that of the omitted
$p_k$ field, so if the original GLSM described a Calabi-Yau manifold, the new GLSM will also satisfy the Calabi-Yau condition. In the gCICY case the charges of the GLSM for ${\cal M}$ sum to $\vec{e}_i$, and the additional field $p_{\vec{e}}$ of charge $-\vec{e}$ restores the Calabi-Yau condition.

So far we have merely developed an alternative presentation of the original $X$.
Next, we describe formerly non-polynomial deformations in this new presentation, where they become polynomial. The relations $P_l$ with $l\neq k$, and $P_{\ve}$ in the gCICY case, are replaced by the most general polynomials in coordinates $(\phi,u)$ of the same multi-degree:
\begin{equation}
P_l(\phi,u)=P_l(\phi)+\sum_{a=0}^{m-1}g_{la}(\phi)\,u_a+\cdots\,,\qquad
P_{\ve}=P_{\ve}(\phi,u)\,,
\label{eq:A-relations}
\end{equation}
with $\deg g_{la}=\vd_l-\vec q$. The ellipsis denotes terms of higher order in the $u_a$ that are allowed by the charges. On the vacuum, where the $u_a$ take the values \eqref{udefs}, these become the rational defining relations. In general, higher-order terms in $u_a$ are not optional. The $u_a$ are the only R-charge-zero fields with negative charge under $U(1)_\sst$, and each carries charge $-1$. So a gCICY relation with $e^{\sst}=-\kappa<0$ is at least of degree $\kappa$ in the $u_a$. Terms linear in $u$ suffice only when, up to field redefinitions, the charges permit nothing else, as in the examples of Sections~2 and~3.

For example, in the case linear in $u_a$, we have explicitly
\begin{equation}
    P_l \: + \: g_{l 0} u_0 + g_{l 1} u_1 + \cdots + g_{l, m-1} u_{m-1},
\end{equation}
where each $g_{l a}$ is a homogeneous polynomial of degree such that $g_{l a} u_a$ has the same degree as $P_l$.
Along the space of vacua, each $u_a$ is equivalent to a fraction, for example,
\begin{equation}
    u_0 \: = \: - \frac{f_0}{x_1},
\end{equation}
so along the vacuum manifold the first term in the deformation of $P_l$ describes a non-polynomial deformation:
\begin{equation}
    g_{l 0} u_0 \: = \: - g_{l 0} \frac{f_0}{x_1}.
\end{equation}
The other terms are similar.
Thus, in the new presentation, the previously non-polynomial deformations are now polynomial.

As a final note, above we have illustrated the replacement of a single defining equation $P_k$ and the appropriate augmentation of fields in the GLSM. In general there can be multiple equations of this type replaced and the procedure is applied to each pair $(\sst,k)$ in turn with a separate set of fields $u_a$ and $\Lambda_i$ for each (this is the case realized in the example given in Section \ref{sect:gcicy}).


\end{document}